\documentclass[prd,preprint,tightenlines,showpacs]{revtex4}

\RequirePackage{amsfonts}
\RequirePackage{amssymb}
\RequirePackage{amsmath}
\RequirePackage{graphicx}
\RequirePackage{makeidx}
\RequirePackage{ifthen}
\RequirePackage{color}
\renewcommand{\a}{\ensuremath{\alpha}}
\renewcommand{\b}{\ensuremath{\beta} }  
\newcommand{\g}{\ensuremath{\gamma}  }
\renewcommand{\d}{\ensuremath{\delta}  }
\newcommand{\e}{\ensuremath{\epsilon}}
\newcommand{\ve}{\ensuremath{\varepsilon}}

\renewcommand{\th}{\ensuremath{\theta} }

\renewcommand{\l}{\ensuremath{\lambda} }

\newcommand{\s}{\ensuremath{\sigma} }

\newcommand{\be}{\begin{eqnarray}}          \newcommand{\ee}{\end{eqnarray}}
\newcommand{\ba}{\begin{align}}          \newcommand{\ea}{\end{align}}
\newcommand{\benum}{\begin{enumerate}}   
\newcommand{\eenum}{\end{enumerate}}
\newcommand{\bitem}{\begin{itemize}}          
\newcommand{\eitem}{\end{itemize}}

\newcommand{\nn}{\nonumber \\ }

\newcommand{\abs}[1]{\ensuremath{\left| #1 \right|}}   

\newcommand{\sci}[2]{\ensuremath{#1\times 10^{#2}}}  
\newcommand{\psci}[2]{\ensuremath{(#1)\times 10^{#2}}}  
 
\newcommand{\fr}[2]{\ensuremath{\frac{#1}{#2}}}

\renewcommand{\matrix}[1]{\begin{pmatrix} #1 \end{pmatrix}}

\definecolor{myblue}{rgb}{.1,.1,.7}
\definecolor{dcyan}{rgb}{.0,.6,.6}
\definecolor{dmagenta}{rgb}{0.6,0.0,0.6}
\definecolor{brown}{rgb}{0.6,0.2,0.}
\definecolor{darkblue}{rgb}{0,0,0.6}
\definecolor{darkred}{rgb}{0.75,0.0,0.0}
\definecolor{darkgreen}{rgb}{0.0,0.6,0.0}

\newcommand{\blue}{\color{blue}}

\newcommand{\usemarker}{Y}
\newcommand{\marker}[1]{
       \ifthenelse{\equal{\usemarker}{Y}}
                     {\mbox{}\marginpar{\tt #1}}{}
               }

\newcommand{\mk}[1]{
\ifthenelse{\equal{\usemarker}{Y}}
{\noindent\hskip -1truecm {\bf\blue$^{#1}$}}{}
}

\newcommand{\myaddress}{Ludwig-Maximilians-University Munich, \\
Sektion Physik. Theresienstr. 37, D-80333. Munich, Germany}

\begin{document}
\title{Textures and hierarchies in quark mass matrices 
\\ with four texture-zeros}
\author{Yu-Feng Zhou}
\affiliation{\myaddress}
\email[Email:]{zhou@theorie.physik.uni-muenchen.de}
\date{\today}
\begin{abstract}
  A systematic analysis on the phenomenologically allowed textures in Hermitian
  quark mass matrices with four texture-zeros is presented. The current data
  allow a full determination of the mass matrix elements, provided that the
  actual number of the parameters in the matrices is less than or equal to ten.
  Besides the standard texture with zeros located in (1,1) and (1,3) positions,
  we find four new type of parallel mass matrices. All of them are found to have
  similar hierarchical structures: three of the nonzero matrix elements are at
  the same order of magnitude with a small geometric hierarchy, while the
  remaining one is relatively much smaller.  These textures show a possibility
  that a realistic quark mass matrix may have an approximate flavor
  $S(2)_{L}\otimes S(2)_{R}$ symmetry, i.e. a partial flavor democracy. The
  absolute values of the CP violating phases in the matrix elements are found to
  be either close to their maximum ($\approx 90^\circ$) or very small ($\leq
  10^{\circ}$).  We also find twenty-six quasi-parallel and seventeen
  non-parallel textures allowed by the experimental data.
\end{abstract}
\preprint{LMU-03-21}
\maketitle

\section{Introduction}
Although the gauge sector  of the standard model (SM) with
$SU(3)_{C}\otimes SU(2)_{L}\otimes U(1)_{Y}$ symmetry is a great success.
The Yukawa sector of the SM is still poorly understood.  The
outstanding problems such as the origins of the fermion masses, the mixing angles
as well as the CP violation remain to be the focus of the  particle
physics. There have been extensive studies on the possible underlying symmetries
in the  Yukawa coupling matrices in the SM, SUSY or other models
( see, e.g.\cite{
Fritzsch:1977za,Fritzsch:1978vd,Fritzsch:1979zq,Stech:1983zc,Fritzsch:1990qm,%
Fritzsch:1995nx,Gill:1995pn,Kuo:1999dt,Branco:1990fj,%
Dimopoulos:1992yz,Dimopoulos:1992za,Hall:1993ni,%
Ramond:1993kv,Anderson:1994fe,Babu:1995kb,Roberts:2001zy}). 
Lacking of a more fundamental theory of the basic interactions at
present, the phenomenological model-independent approaches to searching for
possible textures or symmetries  in  the fermion mass  matrices are still playing  important
roles.

In the SM, the mass term in the Lagrangian is given by
\begin{align}
-\mathcal{L}_{M}=\bar{u}_{_{L}}M^{u} u_{_{R}}+\bar{d}_{_{L}}M^{d} d_{_{R}},
\end{align}
where the mass matrices $M^{u}$ and $M^{d}$ are  three-dimensional complex
matrices. In the most general case, they  contain 36 real parameters, and are
diagonalized through the following bi-unitary transformations
\begin{align}
R^{u\dagger}_{L} M^{u} R^{u}_{R}&=\mbox{diag}\{m_{u},m_{c},m_{t}\},
\nn
R^{d\dagger}_{L} M^{d} R^{d}_{R}&=\mbox{diag}\{m_{d},m_{s},m_{b}\},
\end{align}
with $R^{u,d}_{L,R}$ being the unitary rotation matrices for the left- and right-handed
quark fields. The Cabibbo-Kobayashi-Maskawa (CKM) matrix $V_{CKM}$ is then defined as the
 product of the left-handed rotation matrices for up and down quarks
\begin{align}
V_{CKM}\equiv R^{u\dagger} \, R^{d} .
\end{align}
%

The observed large hierarchies in both the quark masses and the CKM matrix
elements strongly imply  that they may be related to each other through  simple
patterns or textures of the quark mass matrices. As it was noticed long time
ago, the Cabibbo angle $\th_{C}$ can be expressed in terms of quark mass ratio
as $\th_{C}\approx\sqrt{m_{d}/m_{s}}$ \cite{Gatto:1968ss,Cabibbo:1968vn,Oakes:1969vm}, which has
aroused large amount of theoretical attempts in  relating the CKM matrix elements
to the quark mass ratios  through certain textures of the quark mass matrices
\cite{Fritzsch:1977za,Fritzsch:1978vd,Fritzsch:1979zq,Stech:1983zc,Fritzsch:1990qm,%
  Fritzsch:1995nx,Gill:1995pn,Kuo:1999dt,Branco:1990fj}. 
Among the theoretical activities, the models based on the texture-zeros have
achieved some extent of success in phenomenology. A well known example in the
two family case is a model with  zeros in the (1,1) position for both up and down quark
mass matrices, which correctly leads to the above mentioned value of the Cabibbo
angle\cite{Fritzsch:1977za}.  In the three family case, a straight forward
extension is a  model with six texture-zeros in
total, often being  referred to as Fritzsch model\cite{Fritzsch:1978vd,Fritzsch:1979zq}. 
In this model, the zeros are located in (1,1),
(2,2) and (1,3) positions. This model can not only reproduce the  same predictions for the
Cabibbo angle but also  give more relations to the other CKM matrix elements, such as
$|V_{cb}|\approx |\sqrt{m_{s}/m_{b}}-\sqrt{m_{c}/m_{t}}e^{i \phi}| $ and
$|V_{ub}/V_{cb}|\approx m_{u}/m_{c}$ etc..  The same texture was also applied to the
multi-Higgs-doublet models and results in  the widely used Cheng-Sher ansatz
\cite{cheng:1987rs} (for recent discussions, see,
e.g.\cite{Zhou:2001ew,Wu:2001vq}). Other kinds of six-texture-zero pattern were
also used in the $SU(5)$ grand unification theories to improve the predictions
of mass relations between light fermions\cite{Georgi:1979df}.

However,  with the awareness  of the possible  heavy top quark mass, the six-texture-zero model
was found to be disfavored. Phenomenological studies had indicated that the
Fritzsch model could not accommodate the large top quark mass of $m_{t}=175$GeV
and small CKM matrix element of $|V_{cb}|\approx m_{s}/m_{b}\simeq 0.04$
simultaneously\cite{Harari:1987ex}. 
The current data actually rule out all the textures with six
texture-zeros at electroweak  scale\cite{Ramond:1993kv}.  If the up and down quark
mass matrices are allowed to have different numbers of texture-zeros, the
maximum number of zeros are five, and there are only five allowed
patterns\cite{Ramond:1993kv}.  On the other hand, by requiring the up and down
quark mass matrices have parallel structures,  which is more symmetric and natural,
one arrives at the models with
four texture-zeros\cite{Froggatt:1979nt}. A four zeros texture with zeros
located in (1,1) and (1,3) positions has been extensively discussed in the SM in  the
recent years( see, e.g.
\cite{Fritzsch:1995nx,Gill:1995pn,Gill:1997hm,Rasin:1997pn,Chiu:2000gw,Rosenfeld:2001sc}).
There are also attempts to use this texture in the $SO(10)$ grand unification
theories \cite{Chou:1996di,Chou:1997nx}. Recently, it has also been discussed in
the general two-Higgs-doublet model, as it  can lead to stronger suppressions of
the flavor changing neutral currents associating with heavy fermions
\cite{Zhou:2003kd}.

Compared with the textures with five or six texture-zeros, the models with four
zeros has less predictability.  At  first glance,  Hermitian mass matrices
with four texture-zeros in (1,1) and (1,3) positions have 10 free parameters in
total and can not give any prediction.  It is not true if the hierarchy in the
nonzero matrix elements can be obtained from symmetric considerations.  For instance,
the models with  horizontal  $U(2)$ symmetry
\cite{pomarol:1996xc,Caravaglios:2002br} or $D(3n^{2})$ dihedral
symmetry\cite{Chou:1996di} will make the four-texture-zero textures have clear
hierarchies in the remaining nonzero matrix elements. In those cases, the 
four-texture-zero model could be very predictive.  With a hierarchy of $M^{q}(3,3)
\gg M^{q}(2,2)\approx M^{q}(2,3) \gg M^{q}(1,2)$, such  textures can give
roughly correct predictions of $\abs{V_{us}}$ and $\abs{V_{cb}}$, etc..

Note that the current low energy data contain six quark masses  and four
independent CKM matrix elements.  All of them are in reasonable precisions. 
Making use of
these data, it is possible to determine $all$ the nonzero matrix element in the
four-texture-zero textures, provided that the actual number of free parameters is equal to
or less than ten.  This ``bottom-up'' approach, namely, constructing  the possible textures
of quark mass matrices from the low energy data
has been applied to searching  for
the valid five zero textures before\cite{Ramond:1993kv}.  By constructing all
the phenomenologically allowed textures from the data, one may find important
hints of the underlying symmetries in the Yukawa sector.  This approach
is also suitable for  discussing  all the possible textures and hierarchies in
the four-texture-zero models.

Currently, only one possibility of the four-texture-zero model has been carefully
examined in the literature, namely, the texture-zeros located in (1,1) and
(1,3) positions. This texture can be regarded as a simple extension of the
Fritzsch model with (2,2) elements being nonzero.  
It respects the chiral evolution of quark masses  and  
leads to a very useful   parameterization of the CKM matrix \cite{Fritzsch:1997fw,Fritzsch:1998st}. 
However, there is no any
theoretical or experimental justification that this is the only valid texture
with four-texture-zero.  It is very likely that there exist much more valid ones
with the four-texture-zero located in different places  and  with different hierarchies in
the nonzero matrix elements. Those textures, once  being constructed, might
serve as  a useful guide for  model building.

The purpose of this paper is therefore to present a systematic  analysis of  the
phenomenologically allowed textures of Hermitian quark mass matrices with four
texture-zeros.   It  is organized as follows. In section {\bf \ref{c2}}, we
discuss the hierarchy in the standard four-texture-zero model with zeros
located in (1,1) and (1,3) positions. we show that  the nonzero elements of
the quark mass matrices  can be directly determined from the experimental  data with a 
reasonable precision. 
In section {\bf\ref{c3}}, we discuss  all the four-texture-zero models  with parallel
structures, and find four new textures with zeros in other locations.
In section {\bf\ref{c4}}, the valid four-texture-zero models  with quasi-parallel
and non-parallel structure are discussed. We finally conclude in section {\bf\ref{c5}}.


\section{Mass matrices with texture zeros in (1,1) and (1,3) positions }\label{c2}
In general, the three dimensional up and down quark mass matrices have totally 36 parameters,  
much more than the number of the observables.  However, not all
of them are physical.  The ambiguities in the definition of the quark fields in
the flavor basis allow the following unitary transformations which leave
the physics unchanged
\begin{align}
u_{_{L}}&\to U_{L} u_{_{L}},     \quad   d_{_{L}}\to U_{L} d_{_{L}} ,
\nn 
u_{_{R}}&\to V^{u}_{R} u_{_{R}},  \quad  d_{_{R}}\to V^{d}_{R} d_{_{R}} ,
\end{align}
where $U_{L}$, $V^{u(d)}_{R}$ are all unitary matrices.
The quark mass matrices transform accordingly as
\begin{align}
M^{u}&\to U^{\dagger}_{L} M^{u} V^{u}_{R},   \quad M^{d}\to U^{\dagger}_{L}  M^{d} V^{d}_{R}.
\end{align}
Without a loss of generality, one can arrive at physically equivalent mass
matrices with less parameters by  suitable transformations. For instance, by
suitably choosing the unitary matrices $V^{u(d)}_{R}$ and $U_{L}$, the mass
matrices $M^{u(d)}$ can always be arranged to be Hermitian\cite{Frampton:1985qk}. 
Thus in all the bellow discussions, we shall  focus only on the 
textures of Hermitian mass matrices.

By assuming that the  matrix elements in (1,1) and (1,3) position are both vanishing, one arrives at
the following Hermitian mass matrices with four texture-zeros, which is referred to 
as texture P1
\cite{Fritzsch:1995nx,Gill:1995pn,Gill:1997hm,Rasin:1997pn,Chiu:2000gw,Rosenfeld:2001sc}
\begin{align}\label{fourzero}
\intertext{\centerline{Texture P1}}
M_{q}&=\chi_{q}\matrix{
0 & D_{q}e^{i \phi_{D_{q}}} & 0 \\
D_{q}e^{-i \phi_{D_{q}}} & C_{q} & B_{q}e^{i \phi_{B_{q}}} \\
0 & B_{q}e^{-i \phi_{B_{q}}} & A_{q}
}, \qquad (q=u,d) ,
\end{align}
where $A_{q}$, $B_{q}$, $C_{q}$, $D_{q}$ are real parameters 
$\chi_{q}$ is a scaling factor which is  just the mass of the third family quarks,
i.e. $\chi_{u(d)}=m_{t(b)}$.  This texture can be rewritten  as
\begin{align}
M_{q}&= \chi_{q}
\matrix{
1 & 0 & 0 \\
0 &e^{-i\phi_{D_{q}}} &0 \\
0 & 0 &e^{-i(\phi_{B_{q}}+\phi_{D_{q}})}
}
\matrix{
0   & D_{q} &  0 \\
D_{q} & C_{q}   & B_{q} \\
0   & B_{q} &  A_{q}
}
\matrix{
1 & 0 & 0 \\
0 &e^{i\phi_{D_{q}}} &0 \\
0 & 0 &e^{i(\phi_{B_{q}}+\phi_{D_{q}})}
} 
\\
&\equiv V^{\dagger}_{q} \widetilde{M}_{q} V_{q} ,
\end{align}
with $\widetilde{M}_{q}$ being a real symmetric matrix.

For Hermitian quark mass matrices, one can further perform a common unitary
transformation on both left-   and right-handed  quark fields,
which   keeps the mass matrices to be Hermitian and has no
physical effect, namely,
the physical observables  are unchanged under the transformation of 
\begin{align}\label{common-rotation}
M_{q}\to U^{\dagger} M_{q} U \qquad (q=u,d)  .
\end{align}
Taking $U=V_{u}^{\dagger}$, one finds that the up quark mass matrix can alway
be rotated  to be real, and only the phase differences between up and down quark
mass matrix elements are physically relevant. It is then helpful to  define 
two independent  phases
\begin{align}
\phi_{1}\equiv \phi_{D_{u}}-\phi_{D_{d}}, \quad  \mbox{and}\quad
\phi_{2}\equiv \phi_{B_{u}}-\phi_{B_{d}}.
\end{align}
Thus this type of mass matrix actually contains the following ten free parameters
$$
A_{u(d)}, \ B_{u(d)}, \ C_{u(d)}, \ D_{u(d)}, \ \phi_{1}, \ \mbox{and }\  \phi_{2}.
$$
The eigenvalues of the matrix $\widetilde{M}_{q}$ is denoted by 
$\l_{1q},\l_{2q}$ and $\l_{3q}$, which are just the same as  the quark masses  $m_{iq}$ up to a 
sign difference.  As a phase convention, we require $\l_{3q}>0$.
Since det$(\widetilde{M}_{q})=-A_{q} D_{q}^2$,  the value of $\l_{1q}$ and $\l_{2q}$
must have different signs. We then define $\l_{1q}=-\eta \, m_{1q}$ and $\l_{2q}=\eta \,m_{2q}$
with $\eta=\pm 1$ to distinguish the two cases.
Making use of the  transformation invariants such as
$\mbox{Tr}(\widetilde{M})$, $\mbox{Tr}(\widetilde{M}^{2})$ and 
$\mbox{det}(\widetilde{M})$, one finds that the value of $C_{q}$,
$D_{q}$ and  $B_{q}$ depend only on the value of $A_{q}$
%
\begin{align}
C_{q}&=\l_{1q}+\l_{2q}+\l_{3q} -A_{q} ,\nn
D_{q}&=\sqrt{-\fr{\l_{1q}\l_{2q}\l_{3q}}{A_{q}}} , \nn
B_{q}&=\sqrt{-\fr{(A_{q}-\l_{1q})(A_{q}-\l_{2q})(A_{q}-\l_{3q})}{A_{q}}} .
\end{align}
Thus the hierarchical
structure of this type of  mass matrix is completely determined by $A_{q}$.

The rotation matrix $\widetilde{R}_{q}$ which diagonalize $\widetilde{M}_{q}$
  can be obtained by   solving the characteristic equation
and is found to be   
%
%
\begin{align}
\widetilde{R}_{q}\simeq
\matrix{
1 & \eta \ve_{q}  & \ve^{3/2}_{q} \d_{q} \\
-\eta \ve_{q}(1-\fr12 \d^2_{q}) & 1- \fr12 \d^{2}_{q} & -\d_{q}\\
\eta \ve_{q}\d_{q} & -\d_{q} & 1-\fr12 \d^{2}_{q}
} ,
\end{align}
where we have used $|\l_{3q}|, A_q \gg|\l_{2q}|\gg|\l_{1q}|$ 
and the well known  approximate relation of $|\l_{1q}/\l_{2q}|\simeq
|\l_{2q}/\l_{3q}|$.  One sees that the rotation matrix  depends only on two variables $\ve_{q}$ and $\d_{q}$,
which are defined through
\begin{align}
\ve^2_{q}\equiv -\fr{\l_{1q}}{\l_{2q}} \quad \mbox{and} \quad \d_{q}^{2}\equiv\fr{\l_{3q}-A_{q}}{\l_{3q}} .
\end{align}
The CKM matrix is  given by
\begin{align}
V_{CKM}&=R^{\dagger}_{u}R_{d}
\nn
&=\widetilde{R}^{T}_{u}V_{u}V_{d}^{\dagger} \widetilde{R}_{d}
\nn
&=\widetilde{R}^{T}_{u}
\matrix{
1 & 0 & 0 \\
0 & e^{i \phi_{1}} & 0 \\
0 &  0 & e^{i(\phi_{1}+\phi_{2})}
}
\widetilde{R}_{d} .
\end{align}

The absolute values of some of the CKM matrix elements are therefore given as follows (for
$\eta=\pm 1$):
\begin{align}
\abs{V_{us}}&\simeq\abs{
\ve_{d}- \ve_{u}\left(1-\fr{\d_{d}^{2}}{2}\right)\left(1-\fr{\d_{u}^{2}}{2}\right)e^{i \phi_{1}}-\d_{u}\d_{d}\ve_{u}e^{i (\phi_{1}+\phi_{2})}
}, \label{Vus}
\\
\abs{V_{cb}}&\simeq\abs{
\d_{d}\left(1-\fr{\d_{u}^{2}}{2}\right)e^{i \phi_{1}}-\d_{u}\left(1-\fr{\d_{d}^{2}}{2}\right)e^{i (\phi_{1}+\phi_{2})} \pm\d_{d}\ve_{u}\ve_{d}^{3/2}
},
\\
\abs{V_{ub}}&\simeq\abs{
\ve_{u}\left[\d_{d}\left(1-\fr{\d_{u}^{2}}{2}\right)e^{i \phi_{1}}-\d_{u}\left(1-\fr{\d_{d}^{2}}{2}\right)e^{i (\phi_{1}+\phi_{2})}   
\right]\mp\d_{d}\ve_{d}^{3/2}
},
\\
\abs{V_{ts}}&\simeq\abs{
\d_{d}\left(1-\fr{\d_{u}^{2}}{2}\right)e^{i \phi_{1}}-\d_{u}\left(1-\fr{\d_{d}^{2}}{2}\right)e^{i (\phi_{1}+\phi_{2})} \mp\d_{u}\ve_{d}\ve_{u}^{3/2}
}.
\end{align}
In a good approximation, the four zeros texture can reproduce the well known relation of
\begin{align}
\abs{V_{us}}\simeq\abs{\ve_{d}- \ve_{u} e^{i \phi_{1}}}
\quad \mbox{and} \quad
\abs{V_{cb}}\simeq \abs{V_{ts}}=\abs{\d_{d}-\d_{u}e^{i \phi_{2}} }.
\end{align}
But the ratio of $\abs{ V_{ub}/V_{cb}}$ may deviate from  $\sqrt{m_{u}/m_{c}}$,
 as the  corrections from the last terms in the expressions of $|V_{ub}|$
and $|V_{cb}|$ can  not be ignored, which  makes it to be able to accommodate the current data of 
$\abs{ V_{ub}/V_{cb}}$  that  is slightly  larger than  $\sqrt{m_{u}/m_{c}}$ \cite{Fritzsch:2002ga}.

The above analytical expressions, although useful in understanding the general
behavior of this type of mass matrices, can not directly determine the hierarchy
in the nonzero matrix elements, since the values of $\d_{q}$ or $A_{q}$ are still
unknown.  A quantitative determination of all the matrix element can only be
done through numerical methods. Since this texture contains only ten free parameters.
It is then possible to fit all  the parameters from the data on six quark
masses and four independent CKM matrix elements.  Through a global fit to the experimental
data, one is able to answer two important questions: 1) Can this texture 
 accommodate all  the current data consistently? 2) What is the exact hierarchy in the matrix
elements?

For the $\chi^{2}$ fit, we take the input parameters  of the quark masses mainly from
Ref.\cite{Hagiwara:2002fs}. But for strange quark masses, we shall adopt the  recent
update of $\bar{m}_{s}(2\mbox{GeV})=117.4\pm 23.4$ MeV from Refs.\cite{Narison:2002hk,Narison:1999uj} which
is a combined average of the strange quark mass from chiral perturbative theory and QCD
spectrum sum rules.  Using the renormalization group equation, we rescale 
all the quark masses to the energy scale of $m_{Z}$, which gives the following values
\begin{align}\label{quark-mass-data}
m_{u}(m_{Z})&=(0.000883-0.00294) \mbox{ GeV} ,
\nn 
m_{c}(m_{Z})&=(0.589-0.691) \mbox{ GeV}  ,
\nn 
m_{t}(m_{Z})&=(178-189) \mbox{ GeV}  ,
\nn 
m_{d}(m_{Z})&=(0.00177-0.0053) \mbox{ GeV} ,
\nn 
m_{s}(m_{Z})&=(0.0553-0.0883) \mbox{ GeV}  ,
\nn 
m_{b}(m_{Z})&=(2.76-3.04) \mbox{ GeV}  .
\end{align} 
We then choose  four independent observables related to the CKM matrix elements.
The values of the CKM matrix element $|V_{us}|$, $|V_{cb}|$ and $|V_{ub}|$ can be
directly obtained from the measurement of semileptonic kaon and $B$ meson decays.
The value of  $|V_{us}|$ and  $|V_{cb}|$ have been measured with a good precision \cite{Hagiwara:2002fs}
\begin{align}\label{Vus-data}
|V_{us}|&=0.2196\pm 0.0023,
\quad
|V_{cb}|=0.04\pm 0.013\pm 0.009 .
\end{align}
The  value of $|V_{ub}|$ is  extracted from semileptonic decays $B\to \pi(\rho) \ell \nu$,
which suffers from large theoretical uncertainties from  the heavy quark effective theory.
The latest data give \cite{Bornheim:2002du,Bigi:1999dv}
\begin{align}\label{Vub-exp-data}
|V_{ub}|=
\left\{
\begin{array}{l}
\psci{4.09\pm 0.59\pm 0.69}{-3}\qquad \mbox{(LEP)},
\nn
\psci{4.08\pm 0.56\pm 0.40}{-3} \qquad \mbox{(CLEO)}.
\end{array}
\right.
\\
\end{align}
There are many  CP violating parameters, for example,  the angle of the unitarity
triangle $\a,\b$ or $ \g$. The angle $\b$ can be  extracted from mixing induced CP
violating processes such as $B\to J/\psi K_{S}$ without hadronic uncertainties,
but there is no perfect method to cleanly extract the  angles $\a$ and $\g$, as
the interference between strong and week phase is complicated. ( for detailed
discussions on the strong and week phases in B decays, see
e.g.
\cite{Gronau:1990ka,Lipkin:1991st,Fleischer:1998um,Neubert:1998pt,He:2000ys,%
Zhou:2000hg,Wu:2000rb,Wu:2002nz}). Here, we use  an  other important
CP violation parameter,  the Jarlskog rephasing
invariant CP violation quantity $J_{CP}$,  defined through
$J_{CP}\equiv\mbox{Im}(V_{ir}V_{js}V^{*}_{is}V^{*}_{jr})$ \cite{Jarlskog:1985ht,Wu:1986ea}.
The value of $J_{CP}$ from the recent global fit of the unitarity triangle is given by \cite{ckm-fitter}
\begin{align}\label{JCP-data}
J_{CP}=\psci{2.727\pm 0.233}{-5}. 
\end{align}
Using the about  data as inputs,  the best fitted values of 
mass matrix elements in the four-texture-zero texture of Eq.(\ref{fourzero}) are found to be 
\begin{align}\label{P1}
\intertext{\centerline{Texture P1}}
M^{u}=&\chi_{u}\begin{pmatrix}
\mathbf{0}&(0.0001977\pm 0.00004.989) &\mathbf{0} \\
\thicksim&(0.07398\pm 0.03128)&(0.2555\pm 0.05371) \\
\thicksim&\thicksim&(0.9295\pm 0.03761)\\
\end{pmatrix}, 
\nn\nn
M^{d}=&\chi_{d}\begin{pmatrix}
\mathbf{0}&(0.005731\pm 0.001515)e^{i(-87.5\pm 36.27)^\circ}&\mathbf{0}\\
\thicksim&(0.1052\pm 0.03922)&(0.2702\pm 0.06171)e^{i(6.915\pm 4.946)^\circ}\\
\thicksim&\thicksim&(0.9184\pm 0.05849)\\
\end{pmatrix} ,
\end{align}
where the symbol ``$\thicksim$''  stands for 
the repeated or known matrix elements, 
since the matrices are Hermitian.
From the fit results, we arrive at the following observations\\
\begin{enumerate}
\item A very consistent fit with the minimum $\chi^{2}$ of
  $\chi^2_{min}=\sci{1.1}{-5}$ is obtained, which clearly indicates that this
  texture is in a remarkable agreement with all the current data. Thus it is a
  good candidate for  realistic quark mass matrices.

\item The value of $A_{q}$ are determined in  a good  precision and
are found to be
\begin{align}
A_{u}=0.9295\pm 0.03761\quad \mbox{and } \quad A_{d}=0.9184\pm 0.05849.
\end{align}
However,  the deviation of $A_{q}$ from unity is obvious. 
Accordingly, the value of
$\d_{q}$s are
\begin{align}
\d_{u}=0.2655\pm0.07083 \quad \mbox{and} \quad \d_{d}=0.2857\pm0.1023, 
\end{align}
which are not very small. With these
values  the mass matrices  in Eq.(\ref{P1})can be written as the  following approximate form 
\begin{align}
M^{u}=\chi_{u}\begin{pmatrix}
\mathbf{0}& \ve_{u}^3 & \mathbf{0}\\
\thicksim& \d_{u}^2& \d_{u}\\
\thicksim&\thicksim&1-\d_{u}^2\\
\end{pmatrix},
\quad
M^{d}=&\chi_{d}\begin{pmatrix}
\mathbf{0}&  \ve_{d}^3 e^{\phi_{1}} & \mathbf{0}\\
\thicksim& \d_{d}^2  & \d_{d} e^{\phi_{2}}\\
\thicksim&\thicksim&1-\d_{d}^2\\
\end{pmatrix} .
\end{align}
This is in strong disagreement with the texture proposed in
Ref.\cite{Ramond:1993kv}, but agrees with the recent qualitative discussion in
Ref.\cite{Fritzsch:2002ga}.  In both up and down quark mass matrices, comparing with the small
(1,2) element of $\sci{5.7}{-3}(\sci{1.9}{-4})$ for down (up) quark mass matrix, the (2,2), (2,3) and (3,3)
elements are significantly larger and are roughly  at the same order of
magnitude.  The ratios between (2,2), (2,3) and (3,3) elements are
\begin{align}
\abs{\fr{M^{q}(2,2)}{M^{q}(2,3)}}\simeq \abs{\fr{M^{q}(2,3)}{M^{q}(3,3)}}\simeq \d_{q}\simeq 0.3 .
\end{align}
%

\item The absolute central value of one phase $\phi_{1}$ is found to be close to
  $90^{\circ}$,  i.e $|\phi_{1}|\simeq 90^{\circ}$, which implies the possibility of the maximum CP violation
  \cite{Wolfenstein:1984ta,Gronau:1985nm,Fritzsch:1995nx}, in the sense that the
  CP violating phase may take it maximum value in the real world.    
However, the
  error of about $40^{\circ}$ is quite large. At present one can not draw a
  robust conclusion on that.  The reason that the $\phi_{1}$ is poorly
  determined can be understood from the expression of $|V_{us}|$ in
  Eq.(\ref{Vus}), in which the $\phi_{1}$ gives  the interference between
  $\ve_{d}$ and $\ve_{u}$. Since they differ by two order of magnitudes, the
  interference is rather weak and can not give enough  information for the value of
  $\phi_{1}$. The other phase parameter $\phi_{2}$ is found to be small ($\leq
  10^{\circ}$) but nonzero.

\end{enumerate}


\section{Four zeros with parallel textures}\label{c3}
 We proceed to discuss other possible textures with four texture-zeros. To this
 end, we first classify all the possible textures into three types, depending on the
 locations of the texture-zeros, namely,
\begin{enumerate}
\item Parallel textures. The texture-zeros have the same locations for
  both up and down quark mass matrices. There are totally 15 different textures
  which can be further divided into three catalogs: 1.a) Three textures with
  zeros both in off-diagonal positions.  1.b) Three textures with  the
  texture-zeros both in diagonal positions.  1.c) Nine
  textures in which one zero is in diagonal position while the other one is
  off-diagonal. 
  
\item Quasi-Parallel textures. For each quark mass matrix there exist two
  texture-zeros, but their locations are different for up and down quark mass
  matrices.  There are 210 different textures in total.
  
\item Non-Parallel textures. One of the up and down mass matrix have three zeros
  while the other one have only one zero. There are 240 possibilities.
\end{enumerate}  
 
Among those textures, the mass matrices with parallel textures are of course the
most simplest and symmetric ones, and therefore are of particular interest.
Note that for the case 1.a of the parallel texture, there are 
9 free parameters, i.e. four matrix elements for each quark mass  matrix
and one relative phase. The number of parameters is less than that of the 
observables. Thus they are over determined. 
%
For case 1.b there are totally 12
free parameters, more than the number of the available data. So, the matrix
elements of these textures remain to be undetermined and will not be discussed
here. For case 1.c, there are 10 parameters in total.  In principle all the
matrix elements can be determined.

We adopt the $\chi^{2}$ fit method to find all the phenomenologically allowed
mass matrices with four texture-zeros and parallel textures. To this end, 
a selection criterion is needed.  In the fits we shall exclude all
the textures which lead to any of the best fitted values of the quark masses,
the CKM matrix elements  $|V_{us}|$, $|V_{ub}|$, $|V_{cb}|$ or the $J_{CP}$
outside the $1\s$ experimentally allowed ranges  in Eqs.(\ref{quark-mass-data}), (\ref{Vus-data}),
(\ref{Vub-exp-data}) and  (\ref{JCP-data}). 
In the parallel textures, by using Eq.(\ref{common-rotation}), one can always rotate the up quark
mass matrix to be real.
We then choose it   as a phase convention.  The fit
results show four phenomenologically allowed textures other than the texture P1,
being referred to as texture P2-P5, These textures as well as the best fitted matrix
elements are as follows:

\vskip 12pt 1) Texture P2: Zeros are located in (1,1) and (1,2) positions. The
best fitted up and down quark mass matrices are

\begin{align}
\intertext{\centerline{Texture P2}}
M^{u}=&\chi_{u}\begin{pmatrix}
\mathbf{0}&\mathbf{0} &(0.000279\pm 0.00007289) \\
\thicksim&(0.467\pm 0.01485)&(0.497\pm 0.006079) \\
\thicksim&\thicksim&(0.5365\pm 0.01672)\\
\end{pmatrix} ,
\nn\nn
M^{d}=&\chi_{d}\begin{pmatrix}
\mathbf{0}&\mathbf{0}&(0.007886\pm 0.0008101)e^{i(87.31\pm 43.98)^\circ}\\
\thicksim&(0.4851\pm 0.03067)&(0.4875\pm 0.02665)e^{i(-175.5\pm 0.3352)^\circ}\\
\thicksim&\thicksim&(0.5385\pm 0.03412)\\
\end{pmatrix} ,
\end{align}
with a minimum $\chi^{2}$ of $\chi^2_{min}=\sci{1.6}{-6}$ which indicates a very
consistent fit.  From the numerical values, one sees that in this texture 
 the elements associating with the second and the third families, i.e. the 
elements in (2,2), (2,3) and (3,3) positions could be  very close to
each other. The ratios between them are given by 
\begin{align}
\abs{\fr{M^{q}(2,2)}{M^{q}(2,3)}} \simeq \abs{\fr{M^{q}(2,3)}{M^{q}(3,3)}} \simeq 1,
\qquad (q=u,d).
\end{align}
Thus in a good approximation this texture  has the following simple form
\begin{align}\label{P2}
M^{q}\simeq\fr{\chi_{q}}{2}
\left[
\matrix{
0 & 0  & 0  \\
0 & 1  & 1 \\
0 & 1  & 1
}
+
\matrix{
0 & 0   &  \mathcal{O}(\ve_{q}^3)  \\
0 & 0   &0 \\
\mathcal{O}(\ve_{q}^3) & 0  & 0
}
\right] , \qquad (q=u,d) .
\end{align}
It was noticed long ago that the large hierarchy in the quark masses suggests that the realistic
quark mass matrices are close to a "democracy" limit \cite{Fritzsch:1979zq,Fritzsch:1990qm} 
\begin{align}
M^{q} \approx \fr{\chi_{q}}{3}\matrix{
1 & 1 & 1 \\
1 & 1 & 1 \\
1 & 1 & 1
}, \qquad (q=u,d),
\end{align}
which poses a flavor $S(3)_{L}\otimes S(3)_{R}$ symmetry. To generate the quark
masses of the first and the second families and also the mixing angles, such a
symmetry must be broken down. A popular way is to break the symmetry through the
steps of $S(3)_{L}\otimes S(3)_{R}\to S(2)_{L}\otimes S(2)_{R} \to \times,$ (
see, e.g. \cite{Fritzsch:1990qm,Fritzsch:1994yx,Kang:1997uv}), where the
$S(2)_{L}\otimes S(2)_{R}$ is associated with the first and the second family
quarks.  However, the texture P2 found here shows an example that the symmetry
of $S(3)_{L}\otimes S(3)_{R}$ may be firstly broken down to a $S(2)_{L}\otimes
S(2)_{R}$ symmetry associating with the second and the third families, and then
the $S(2)_{L}\otimes S(2)_{R}$ symmetry is slightly broken down by a tiny
perturbation to the (1,3) element. The small value of
$M^{q}(1,3)/M^{q}(2,2)\approx \mathcal{O}(10^{-2}\sim 10^{-3})$ indicates
the symmetry breaking strength.  
Thus the realistic
quark mass matrix could  have an approximate $S(2)_{L}\otimes S(2)_{R}$ symmetry,
namely, a partial flavor democracy.   
In this texture, one also find a large
phase of $87^{\circ}$ which is close to the maximum ( $90^{\circ}$ ) in
(1,3) element of the down quark mass matrix.  The possibility of the maximum
phase therefore does not uniquely belong to the texture P1. 
It is of interest to note that starting from such a partial
flavor democracy some interesting relations between the quark masses and the mixing angles can be
correctly reproduced \cite{in-preparation}.

\vskip 12pt  
2) Texture P3: Zeros  are located in (1,2) and (2,2) positions. The best fitted  matrix
elements are found to be 
\begin{align}
\intertext{\centerline{Texture P3}}
M^{u}=&\chi_{u}\begin{pmatrix}
(0.3788\pm 0.03425)&\mathbf{0} &(0.4828\pm 0.01464) \\
\thicksim&\mathbf{0}&(0.0003199\pm 0.00007) \\
\thicksim&\thicksim&(0.6243\pm 0.03724)\\
\end{pmatrix},
\nn\nn
M^{d}=&\chi_{d}\begin{pmatrix}
(0.4244\pm 0.04044)&\mathbf{0}&(0.4829\pm 0.02282)e^{i(-2.756\pm 0.4887)^\circ}\\
\thicksim&\mathbf{0}&(0.006209\pm 0.0005847)e^{i(-176\pm 10.51)^\circ}\\
\thicksim&\thicksim&(0.6042\pm 0.04269)\\
\end{pmatrix}.
\end{align}
with a minimum $\chi^{2}$ of $\chi^2_{min}=0.86$. The numerical values show that
the (1,1) (1,3) and (3,3) elements are at  the same order of magnitude with a small
geometric hierarchy of
\begin{align}
\abs{\fr{M^{q}(1,1)}{M^{q}(1,3)}}\approx\abs{\fr{M^{q}(1,3)}{M^{q}(3,3)}}\approx 0.7, \qquad (q=u,d).
\end{align}
Therefore  it can be roughly understood as another type of  partial flavor democracy between the first and the third families. 
On the other hand, the (2,3) elements are found to be much smaller.  In this texture,
 the quark
masses of the second family arise purely from the mixing terms. A small phase of
a few degrees in (1,3) element and a  phase around $\sim -180^{\circ}$
in  the (2,3) element in down quark mass matrix are found. It shows 
however that the quark mass matrices are nearly  real. Considering the fact
that the experimentally observed effects of CP violation is quite small, this texture 
with small imaginary matrix elements  is also  natural.  

\vskip 12pt  
3) Texture P4. Zeros  in (2,2) and (2,3) elements. The best fitted  
mass matrices are
\begin{align}
\intertext{\centerline{Texture P4}}
M^{u}=&\chi_{u}\begin{pmatrix}
(0.07226\pm 0.01143)&(0.0001978\pm 0.0000486) &(0.2526\pm 0.01951) \\
\thicksim&\mathbf{0}&\mathbf{0} \\
\thicksim&\thicksim&(0.9313\pm 0.03255)\\
\end{pmatrix},
\nn\nn
M^{d}=&\chi_{d}\begin{pmatrix}
(0.103\pm 0.02048)&(0.005712\pm 0.001497)e^{i(88.43\pm 35.07)^\circ}&(0.2669\pm 0.03301)e^{i(8.9\pm 4.354)^\circ}\\
\thicksim&\mathbf{0}&\mathbf{0}\\
\thicksim&\thicksim&(0.9204\pm 0.04911)\\
\end{pmatrix}.
\end{align}
with  $\chi^2_{min}=\sci{4.5}{-4}$.
One sees that   (1,1) (1,3) and (3,3) elements are roughly at the same order of magnitudes with 
a geometric hierarchy of 
\begin{align}
\abs{\fr{M^{q}(1,1)}{M^{q}(1,3)}}\approx \abs{\fr{M^{q}(1,3)}{M^{q}(3,3)}} \approx 0.3, \qquad (q=u,d).
\end{align}
while the (1,2)  element are extremely small. In this texture,  the quark masses of the second family 
also come from mixing terms. The phase of the (1,2) element is large
and close to $90^\circ$.

\vskip 12pt  
4) Texture P5. Zeros  in (2,3) and (3,3) elements. The best fitted up and down quark
mass matrices are given by
\begin{align}
\intertext{\centerline{Texture P5}}
M^{u}=&\chi_{u}\begin{pmatrix}
(0.06263\pm 0.001382)&(0.2489\pm 0.002558) &(0.0001981\pm 0.00002745) \\
\thicksim&(0.934\pm 0.01647)&\mathbf{0} \\
\thicksim&\thicksim&\mathbf{0}\\
\end{pmatrix}
\\
M^{d}=&\chi_{d}\begin{pmatrix}
(0.0980\pm 0.00313)&(0.259\pm 0.00565)e^{i(7.79\pm 1.23)^\circ}&(0.00567\pm 0.000632)e^{i(-90.3\pm 17.3)^\circ}\\
\thicksim&(0.9255\pm 0.02941)&\mathbf{0}\\
\thicksim&\thicksim&\mathbf{0}\\
\end{pmatrix}
\end{align}
with $\chi^2_{min}=\sci{3}{-3}$. Clearly, their is a similar   geometric hierarchy in the 
(1,1),  (1,2) and (2,2) elements:
\begin{align}
\abs{\fr{M^{q}(1,1)}{M^{q}(1,2)}}\approx \abs{\fr{M^{q}(1,2)}{M^{q}(2,2)}} \approx 0.3 .
\end{align}
%
Unlike all the other textures with large (3,3) elements,
 it has zeros in (3,3) element while a large (2,2) element with $M(2,2)\simeq
1$. In this type of mass matrices, the large third family quark masses are purely
generated from small mixing terms. This mismatched texture can not be generated
from the chiral evolutions of the quark mass matrices. One sees again that the 
smallest matrix element in (1,3) position is accompanied with a large phase
with the absolute value around $90^{\circ}$.

In summary, besides the standard four zero textures, there are totally four
phenomenologically allowed quark mass matrices with parallel textures. These
textures, although differ in the location of the texture-zeros, show  similar
hierarchical structures: three of the elements are roughly at order one with a
small geometrical hierarchy, and the other one one is much smaller. It indicates 
that some of  those textures may be described by a partial flavor democracy for the
related families, and the small elements may have a perturbative origin. The 
large CP violating  phases with the absolute value around its maximum ($\sim
90^{\circ}$) are found in texture P1, P2, P4 and P5.  But in texture P3, the
matrix elements are almost real.  
Thus the values of the CP violating phases strongly depend on the locations 
of the texture-zeros.


\section{quasi-parallel and non-parallel textures}\label{c4}
If one abandons the parallel condition, within the framework of the four
texture-zeros, much more phenomenologically allowed textures can be constructed. As it is mentioned
in the previous sections, those textures can be roughly divided into two types:
quasi-parallel and non-parallel textures.

For quasi-parallel textures, the up and down quark mass matrices may have two zeros
in different places. 
If  both of the mass matrices have two zeros  in the off-diagonal positions
there will be 8 parameters left. 
If one of the quarks mass matrices  have two zeros in the off-diagonal positions
while the other one have only one zero in the off-diagonal position, there will be 9
free parameters. In this case,  the mass matrix which has one zero in the off-diagonal position
can be rotated to be real by a suitable unitary transformation from
Eq.(\ref{common-rotation}). 
If both of the mass matrices  have one off-diagonal zero, the number of free
parameters is 10. One can just take the up quark mass matrix to be real as in 
the section \ref{c2}. 
Thus, in all the above cases, the matrix elements can be
determined by the experimental data. 
A direct search shows that there are 26
allowed quasi-parallel  textures.  Those textures  
as well as the best fitted matrix elements are summarized  in  appendix A.

Due to the much larger hierarchy in the up quark masses than that in the down
quark masses, compared with textures in the up quark mass
matrices, the one in  the down quark mass matrix plays a more fundamental  role
in determining the CKM matrix elements.  This down quark dominance partially
explains why there exist  many allowed quasi-parallel textures with four
texture-zeros: for certain type of down quark mass matrix, there could be more
than one type of  phenomenologically allowed up quark mass matrices.  
For instance, the fit results show that there are six other
allowed textures in which the zeros are located in (1,1) and (1,3) positions in down
quark mass matrix but the zeros are in different palaces in the up quark mass
matrix.  The related best fits of the matrix elements are listed in Eqs. (\ref{4zero-1}),
(\ref{4zero-2}),(\ref{4zero-3}),(\ref{4zero-4}),(\ref{4zero-5}),(\ref{4zero-6})
in the appendix A. 
A typical one in Eq.(\ref{4zero-5})  is given by
\begin{align*}
M^{u}=&\chi_{u}\begin{pmatrix}
(-0.00001738\pm 0.00001811)&(0.0003112\pm 0.00009818) &\mathbf{0} \\
\thicksim&\mathbf{0}&(-0.05882\pm 0.002509) \\
\thicksim&\thicksim&(0.9965\pm 0.02998)\\
\end{pmatrix},
\nn
M^{d}=&\chi_{d}\begin{pmatrix}
\mathbf{0}&(0.005366\pm 0.001217)e^{i(93.5\pm 14.31)^\circ}&\mathbf{0}\\
\thicksim&(0.03348\pm 0.005211)&(0.09573\pm 0.009601)e^{i(-177.4\pm 73.63)^\circ}\\
\thicksim&\thicksim&(0.9904\pm 0.04784)\\
\end{pmatrix}.
\end{align*}
In this  texture, although the up quark mass matrices are significantly
different than that in the texture P1, the down quark mass matrices show a
similar hierarchy, and the absolute value of the phase in (1,3) element 
is also close to $90^{\circ}$.

Similarly, there are three more textures (see Eqs.
(\ref{4zeroB-1}),(\ref{4zeroB-2}) and (\ref{4zeroB-3})) in which the zeros are located in  (1,1) and
(1,2) positions  in down quark mass matrices. For example in Eq.(\ref{4zeroB-2}), the 
following mass matrix is allowed
\begin{align*}
M^{u}=&\chi_{u}\begin{pmatrix}
(-0.00001042\pm 0.000005605)&\mathbf{0} &\mathbf{0} \\
\thicksim&(0.7212\pm 0.09639)&(0.4473\pm 0.048) e^{i(-4.518\pm 0.6164)^\circ}\\
\thicksim&\thicksim&(0.2822\pm 0.09429)\\
\end{pmatrix} , 
\nn
M^{d}=&\chi_{d}\begin{pmatrix}
\mathbf{0}&\mathbf{0}&(0.006616\pm 0.001082)\\
\thicksim&(0.697\pm 0.09951)&(0.4672\pm 0.04813)\\
\thicksim&\thicksim&(0.2796\pm 0.104)\\
\end{pmatrix} ,
\end{align*}
where, the partial democratic form  in the down quark mass matrix still remains
to some extent, despite a  different form of the up quark mass matrix. 

For non-parallel textures, i.e., one of the up and down quark mass matrices have
three texture zeros, the maximum number of parameter are usually less than or
equal to 10, except for the ones with totally three zeros in diagonal position.
For completeness, we also
perform a primitive search for the valid textures and find 15 allowed types. 
%
This type of texture is of less interest as it is less symmetric, and has not
yet been considered in the model constructions. 
The details of those allowed textures of this type will not be discussed there.
However,  note   that although the Fritzsch model with six texture zeros in (1,1),  (1,3) and (2,2)
positions in both up and down quark mass matrices have been ruled out, the down
quark mass matrix can have three zeros in such positions alone. 
For example, 
the following quark mass matrix is found to be allowed 
\begin{align*}
M^{u}=&\chi_{u}\begin{pmatrix}
(0.000005803\pm 0.00003106)&\mathbf{0} &(0.001197\pm 0.00138) \\
\thicksim&(0.03942\pm 0.01812)&(0.1858\pm 0.04586) \\
\thicksim&\thicksim&(0.964\pm 0.03157)\\
\end{pmatrix} ,
\nn
M^{d}=&\chi_{d}\begin{pmatrix}
\mathbf{0}&(-0.005556\pm 0.001399)e^{i(95.08\pm 50.67)^\circ}&\mathbf{0}\\
\thicksim&\mathbf{0}&(0.1535\pm 0.01946)e^{i(3.515\pm 45.09)^\circ}\\
\thicksim&\thicksim&(0.9764\pm 0.04894)\\
\end{pmatrix} ,
\end{align*}
with a large phase of $\phi=95^{\circ}$ in the (1,2) elements of the down quark
mass matrix.


\section{conclusions}\label{c5}
In conclusion, we have systematically explored  all the possible four zero
textures with the number of parameters less than or equal to ten. We have found
that there are totally five allowed mass matrices with parallel textures. 
It is shown  that the phenomenologically allowed mass matrices should have
one relatively small matrix elements while three others are at the same order of
magnitude with a small geometric hierarchy. The hierarchies in the mass matrices
can be written in the following approximate form
\begin{align}
\begin{array}{ccccc}
\mbox{P1} & \mbox{P2} & \mbox{P3} &\mbox{P4} &\mbox{P5} \\
\matrix{
0 &  \e_{1} & 0 \\
\e^{*}_{1} &  \l_{1}^2 & \l_{1}\\
0  & \l_{1} & 1
},&
0.5\matrix{
0 &  0 & \e_{2} \\
0 &  \l^{2}_{2} & \l_{2}\\
\e^{*}_{2} & \l_{2} & 1
},
&
0.6 \matrix{
\l^{2}_{3} &  0 & \l_{3} \\
0 &  0 & \e_{3}\\
\l_{3} & \e^{*}_{3} & 1
},
&
\matrix{
\l^{2}_{4} &  \e_{4} & \l_{4} \\
\e^{*}_{4} &  0 & 0\\
\l_{4} & 0  & 1
},
&
\matrix{
\l^{2}_{5} &  \l_{5} & \e_{5} \\
\l_{5 } &  1 & 0\\
\e^{*}_{5} & 0 & 0
},
\end{array}
\end{align}
with 
\begin{align}
\l_{1}\simeq \l_{4}\simeq \l_{5}\simeq 0.3, \quad \l_{2}\simeq 1.0,\quad \mbox{and} \quad \l_{3}\simeq 0.7. 
\end{align}
The value of $\e_{i}$s are much smaller, i.e., $|\e_{i}|\ll |\l_{i}|$. Large
phases in $\e_{i}$ with the absolute values closing to $90^{\circ}$ are found
for texture P1, P2, P4 and P5.  But in texture P3 all the phases are 
found to be small.  We have also found 26 quasi-parallel textures
and 17 non-parallel textures.
A partial flavor democracy is observed in the texture P2,
which implies that an approximate  flavor permutation symmetry of $S(2)_{L}\otimes S(2)_{R}$
could  remain  after the breaking of $S(3)_{L}\otimes S(3)_{R}$.   This  may serve as a guide for the model
buildings.
The textures found there may also be helpful  in  discussing  the   related problems. For
example, it remains to be seen whether those textures can also be applied
to the lepton mass matrixes in the SM and the Yukawa coupling matrices with
multi-Higgs-doublet models. It could also be helpful to search the similar
textures  in the Yukawa sector of various grand unification theories at high 
energy scale.

%


\begin{acknowledgments}
The author is indebted  to H. Fritzsch and Z.Z. Xing for 
carefully reading the manuscript and many useful comments. 
This work is  supported by the Alexander Humboldt Foundation. 
\end{acknowledgments}
\appendix
\small
\section{Phenomenologically allowed mass matrices with quasi-parallel textures} 
\begin{enumerate}
\item\begin{align}\label{4zero-1}
M^{u}=&\chi_{u}\begin{pmatrix}
\mathbf{0}&\mathbf{0} &(0.002992\pm 0.001939) \\
\thicksim&(0.004123\pm 0.004289)&(-0.02518\pm 0.08469) \\
\thicksim&\thicksim&(0.9993\pm 0.0302)\\
\end{pmatrix}
\nn
M^{d}=&\chi_{d}\begin{pmatrix}
\mathbf{0}&(0.005557\pm 0.001228)e^{i(-84.19\pm 61.65)^\circ}&\mathbf{0}\\
\thicksim&(0.02388\pm 0.005598)&(0.02384\pm 0.03518)e^{i(72.52\pm 245.1)^\circ}\\
\thicksim&\thicksim&(0.9994\pm 0.04827)\\
\end{pmatrix}
\end{align}
\item\begin{align}
M^{u}=&\chi_{u}\begin{pmatrix}
\mathbf{0}&\mathbf{0} &(0.002679\pm 0.0007222) \\
\thicksim&(0.005071\pm 0.0003244)&(0.03973\pm 0.002328) \\
\thicksim&\thicksim&(0.9984\pm 0.02993)\\
\end{pmatrix}
\nn
M^{d}=&\chi_{d}\begin{pmatrix}
\mathbf{0}&(0.005389\pm 0.00149)e^{i(-79.17\pm 111.6)^\circ}&(0.006903\pm 0.002963)e^{i(22.01\pm 52.48)^\circ}\\
\thicksim&(0.02361\pm 0.005742)&\mathbf{0}\\
\thicksim&\thicksim&(1\pm 0.04827)\\
\end{pmatrix}
\end{align}
\item\begin{align}
M^{u}=&\chi_{u}\begin{pmatrix}
\mathbf{0}&\mathbf{0} &(-0.00101\pm 0.00008551) \\
\thicksim&(0.03584\pm 0.001073)&(0.1766\pm 0.002824) \\
\thicksim&\thicksim&(0.9676\pm 0.02039)\\
\end{pmatrix}
\nn
M^{d}=&\chi_{d}\begin{pmatrix}
(0.000232\pm 0.000588)&(-0.00503\pm 0.000445)e^{i(-112\pm 17.5)^\circ}&\mathbf{0}\\
\thicksim&\mathbf{0}&(0.154\pm 0.008)e^{i(-10\pm 3.1)^\circ}\\
\thicksim&\thicksim&(0.976\pm 0.0426)\\
\end{pmatrix}
\end{align}
\item\begin{align}
M^{u}=&\chi_{u}\begin{pmatrix}
\mathbf{0}&\mathbf{0} &(0.002473\pm 0.0001303) \\
\thicksim&(0.005087\pm 0.0003234)&(0.0399\pm 0.002323) \\
\thicksim&\thicksim&(0.9987\pm 0.02991)\\
\end{pmatrix}
\nn
M^{d}=&\chi_{d}\begin{pmatrix}
(0.002504\pm 0.0006562)&(0.005346\pm 0.001301)e^{i(123.4\pm 10.61)^\circ}&\mathbf{0}\\
\thicksim&(0.02347\pm 0.005364)&\mathbf{0}\\
\thicksim&\thicksim&(1\pm 0.04828)\\
\end{pmatrix}
\end{align}
\item\begin{align}\label{4zeroB-1}
M^{u}=&\chi_{u}\begin{pmatrix}
\mathbf{0}&(0.0004591\pm 0.00006657) &\mathbf{0} \\
\thicksim&(0.8188\pm 0.0305)&(-0.3843\pm 0.009781) \\
\thicksim&\thicksim&(0.1846\pm 0.01171)\\
\end{pmatrix}
\nn
M^{d}=&\chi_{d}\begin{pmatrix}
\mathbf{0}&\mathbf{0}&(0.005938\pm 0.001072)e^{i(-87.66\pm 11.46)^\circ}\\
\thicksim&(0.8518\pm 0.04308)&(0.3504\pm 0.02167)e^{i(-180\pm 0.01061)^\circ}\\
\thicksim&\thicksim&(0.1718\pm 0.01536)\\
\end{pmatrix}
\end{align}
\item\begin{align}
M^{u}=&\chi_{u}\begin{pmatrix}
\mathbf{0}&(0.0001908\pm 0.00005187) &\mathbf{0} \\
\thicksim&(0.005055\pm 0.0003245)&(-0.03962\pm 0.002329) \\
\thicksim&\thicksim&(0.9984\pm 0.02992)\\
\end{pmatrix}
\nn
M^{d}=&\chi_{d}\begin{pmatrix}
\mathbf{0}&(0.005559\pm 0.00151)e^{i(-96.31\pm 59.68)^\circ}&(0.005524\pm 0.001065)e^{i(21.95\pm 39.94)^\circ}\\
\thicksim&(-0.02351\pm 0.005726)&\mathbf{0}\\
\thicksim&\thicksim&(0.9999\pm 0.04827)\\
\end{pmatrix}
\end{align}
\item\begin{align}
M^{u}=&\chi_{u}\begin{pmatrix}
\mathbf{0}&(0.05503\pm 0.002538) &\mathbf{0} \\
\thicksim&(0.9958\pm 0.02998)&(-0.02169\pm 0.004031) \\
\thicksim&\thicksim&(-0.00001203\pm 0.000006505)\\
\end{pmatrix}
\nn
M^{d}=&\chi_{d}\begin{pmatrix}
(0.0150\pm 0.00606)&(0.094\pm 0.00537)e^{i(-180\pm 2.37)^\circ}&\mathbf{0}\\
\thicksim&(-0.993\pm 0.0479)&(0.0385\pm 0.00745)e^{i(-5.23\pm 2.8)^\circ}\\
\thicksim&\thicksim&\mathbf{0}\\
\end{pmatrix}
\end{align}
\item\begin{align}\label{4zero-2}
M^{u}=&\chi_{u}\begin{pmatrix}
\mathbf{0}&(-0.00002781\pm 0.00009973) &(0.003193\pm 0.001108) \\
\thicksim&(0.003487\pm 0.0002779)&\mathbf{0} \\
\thicksim&\thicksim&(1\pm 0.02997)\\
\end{pmatrix}
\nn
M^{d}=&\chi_{d}\begin{pmatrix}
\mathbf{0}&(0.005497\pm 0.001175)e^{i(67.18\pm 71.37)^\circ}&\mathbf{0}\\
\thicksim&(0.02511\pm 0.00494)&(0.03905\pm 0.002748)e^{i(-180\pm 59.59)^\circ}\\
\thicksim&\thicksim&(0.9984\pm 0.0482)\\
\end{pmatrix}
\end{align}
\item\begin{align}
M^{u}=&\chi_{u}\begin{pmatrix}
\mathbf{0}&(0.0003036\pm 0.00002996) &(-0.05884\pm 0.002505) \\
\thicksim&(-0.00003703\pm 0.000006764)&\mathbf{0} \\
\thicksim&\thicksim&(0.9965\pm 0.02997)\\
\end{pmatrix}
\nn
M^{d}=&\chi_{d}\begin{pmatrix}
(0.0206\pm 0.00449)&(-0.00603\pm 0.00127)e^{i(50\pm 10.4)^\circ}&(0.0184\pm 0.00327)e^{i(-180\pm 37)^\circ}\\
\thicksim&\mathbf{0}&\mathbf{0}\\
\thicksim&\thicksim&(0.999\pm 0.0483)\\
\end{pmatrix}
\end{align}
\item\begin{align}\label{4zeroB-2}
M^{u}=&\chi_{u}\begin{pmatrix}
(-0.00001042\pm 0.000005605)&\mathbf{0} &\mathbf{0} \\
\thicksim&(0.7212\pm 0.09639)&(0.4473\pm 0.048) e^{i(-4.518\pm 0.6164)^\circ}\\
\thicksim&\thicksim&(0.2822\pm 0.09429)\\
\end{pmatrix}
\nn
M^{d}=&\chi_{d}\begin{pmatrix}
\mathbf{0}&\mathbf{0}&(0.006616\pm 0.001082)\\
\thicksim&(0.697\pm 0.09951)&(0.4672\pm 0.04813)\\
\thicksim&\thicksim&(0.2796\pm 0.104)\\
\end{pmatrix}
\end{align}
\item\begin{align}\label{4zero-3}
M^{u}=&\chi_{u}\begin{pmatrix}
(0.00001042\pm 0.000005605)&\mathbf{0} &\mathbf{0} \\
\thicksim&(0.2833\pm 0.06691)&(0.4478\pm 0.03507) e^{i(-4.513\pm 0.5086)^\circ}\\
\thicksim&\thicksim&(0.7202\pm 0.06983)\\
\end{pmatrix}
\nn
M^{d}=&\chi_{d}\begin{pmatrix}
\mathbf{0}&(0.006604\pm 0.0007225)&\mathbf{0}\\
\thicksim&(0.3135\pm 0.07092)&(0.4463\pm 0.04499)\\
\thicksim&\thicksim&(0.7098\pm 0.06513)\\
\end{pmatrix}
\end{align}
\item\begin{align}
M^{u}=&\chi_{u}\begin{pmatrix}
(0.00001042\pm 0.000005605)&\mathbf{0} &\mathbf{0} \\
\thicksim&(0.005081\pm 0.0003235)&(0.03982\pm 0.00232) e^{i(-117.8\pm 11.23)^\circ}\\
\thicksim&\thicksim&(0.9984\pm 0.02993)\\
\end{pmatrix}
\nn
M^{d}=&\chi_{d}\begin{pmatrix}
\mathbf{0}&(0.005536\pm 0.001156)&(0.0036\pm 0.0002005)\\
\thicksim&(0.02334\pm 0.004875)&\mathbf{0}\\
\thicksim&\thicksim&(1\pm 0.04827)\\
\end{pmatrix}
\end{align}
\item\begin{align}
M^{u}=&\chi_{u}\begin{pmatrix}
(-0.001891\pm 0.0003225)&\mathbf{0} &(-0.03989\pm 0.00231) \\
\thicksim&\mathbf{0}&(-0.004382\pm 0.001207) \\
\thicksim&\thicksim&(0.9984\pm 0.02992)\\
\end{pmatrix}
\nn
M^{d}=&\chi_{d}\begin{pmatrix}
(0.0235\pm 0.0057)&(0.00549\pm 0.00149)e^{i(-87.6\pm 64.1)^\circ}&\mathbf{0}\\
\thicksim&\mathbf{0}&(0.00153\pm 0.00207)e^{i(-46.7\pm 136.9)^\circ}\\
\thicksim&\thicksim&(1\pm 0.0483)\\
\end{pmatrix}
\end{align}
\item\begin{align}
M^{u}=&\chi_{u}\begin{pmatrix}
(0.004253\pm 0.004838)&\mathbf{0} &(-0.02764\pm 0.08707) \\
\thicksim&\mathbf{0}&(0.002949\pm 0.002012) \\
\thicksim&\thicksim&(0.9992\pm 0.03032)\\
\end{pmatrix}
\nn
M^{d}=&\chi_{d}\begin{pmatrix}
(0.0239\pm 0.00573)&(0.00556\pm 0.00124)e^{i(84.8\pm 62.4)^\circ}&(0.0241\pm 0.0373)e^{i(80.4\pm 180)^\circ}\\
\thicksim&\mathbf{0}&\mathbf{0}\\
\thicksim&\thicksim&(0.999\pm 0.0483)\\
\end{pmatrix}
\end{align}
\item\begin{align}\label{4zero-4}
M^{u}=&\chi_{u}\begin{pmatrix}
(0.00000336\pm 0.00000569)&\mathbf{0} &(0.00371\pm 0.000163) e^{i(59.2\pm 11.0)^\circ}\\
\thicksim&(0.00349\pm 0.000278)&\mathbf{0} \\
\thicksim&\thicksim&(1\pm 0.03)\\
\end{pmatrix}
\nn
M^{d}=&\chi_{d}\begin{pmatrix}
\mathbf{0}&(0.005547\pm 0.001157)&\mathbf{0}\\
\thicksim&(0.02495\pm 0.004871)&(0.03903\pm 0.002749)\\
\thicksim&\thicksim&(0.9985\pm 0.0482)\\
\end{pmatrix}
\end{align}
\item\begin{align}
M^{u}=&\chi_{u}\begin{pmatrix}
(0.2771\pm 0.1022)&\mathbf{0} &(0.4504\pm 0.0518) e^{i(4.519\pm 0.6441)^\circ}\\
\thicksim&(0.00001042\pm 0.000005605)&\mathbf{0} \\
\thicksim&\thicksim&(0.7195\pm 0.1041)\\
\end{pmatrix}
\nn
M^{d}=&\chi_{d}\begin{pmatrix}
(0.2794\pm 0.1118)&(0.006618\pm 0.001157)&(0.4671\pm 0.05117)\\
\thicksim&\mathbf{0}&\mathbf{0}\\
\thicksim&\thicksim&(0.6971\pm 0.1063)\\
\end{pmatrix}
\end{align}
\item\begin{align}
M^{u}=&\chi_{u}\begin{pmatrix}
(-0.003488\pm 0.0002779)&\mathbf{0} &(-0.00003017\pm 0.0001379) \\
\thicksim&(1\pm 0.02997)&(0.003267\pm 0.0007881) \\
\thicksim&\thicksim&\mathbf{0}\\
\end{pmatrix}
\nn
M^{d}=&\chi_{d}\begin{pmatrix}
(-0.0219\pm 0.00552)&(0.0409\pm 0.00282)e^{i(50.4\pm 308.7)^\circ}&(-0.0055\pm 0.00143)e^{i(109.4\pm 268.5)^\circ}\\
\thicksim&(0.9984\pm 0.0482)&\mathbf{0}\\
\thicksim&\thicksim&\mathbf{0}\\
\end{pmatrix}
\end{align}
\item\begin{align}\label{4zero-5}
M^{u}=&\chi_{u}\begin{pmatrix}
(-0.00001738\pm 0.00001811)&(0.0003112\pm 0.00009818) &\mathbf{0} \\
\thicksim&\mathbf{0}&(-0.05882\pm 0.002509) \\
\thicksim&\thicksim&(0.9965\pm 0.02998)\\
\end{pmatrix}
\nn
M^{d}=&\chi_{d}\begin{pmatrix}
\mathbf{0}&(0.005366\pm 0.001217)e^{i(93.5\pm 14.31)^\circ}&\mathbf{0}\\
\thicksim&(0.03348\pm 0.005211)&(0.09573\pm 0.009601)e^{i(-177.4\pm 73.63)^\circ}\\
\thicksim&\thicksim&(0.9904\pm 0.04784)\\
\end{pmatrix}
\end{align}
\item\begin{align}
M^{u}=&\chi_{u}\begin{pmatrix}
(0.003488\pm 0.0002664)&(-0.00002778\pm 0.00002023) &\mathbf{0} \\
\thicksim&\mathbf{0}&(0.003194\pm 0.0002309) \\
\thicksim&\thicksim&(1\pm 0.02734)\\
\end{pmatrix}
\nn
M^{d}=&\chi_{d}\begin{pmatrix}
(0.0251\pm 0.00109)&(0.0055\pm 0.000254)e^{i(-67.2\pm 111.2)^\circ}&(0.0391\pm 0.00262)e^{i(-180\pm 112.1)^\circ}\\
\thicksim&\mathbf{0}&\mathbf{0}\\
\thicksim&\thicksim&(0.998\pm 0.0419)\\
\end{pmatrix}
\end{align}
\item\begin{align}\label{4zero-6}
M^{u}=&\chi_{u}\begin{pmatrix}
(0.00001761\pm 0.000006948)&(0.0003119\pm 0.00003149) e^{i(-88.95\pm 12.3)^\circ}&\mathbf{0} \\
\thicksim&(0.00346\pm 0.0002757)&\mathbf{0} \\
\thicksim&\thicksim&(1\pm 0.02997)\\
\end{pmatrix}
\nn
M^{d}=&\chi_{d}\begin{pmatrix}
\mathbf{0}&(-0.005197\pm 0.001161)&\mathbf{0}\\
\thicksim&(0.02587\pm 0.00507)&(0.03907\pm 0.002734)\\
\thicksim&\thicksim&(0.9984\pm 0.0482)\\
\end{pmatrix}
\end{align}
\item\begin{align}
M^{u}=&\chi_{u}\begin{pmatrix}
(0.003457\pm 0.0002755)&(0.0003266\pm 0.00003135) e^{i(-46.19\pm 10.07)^\circ}&\mathbf{0} \\
\thicksim&(0.00002034\pm 0.000006928)&\mathbf{0} \\
\thicksim&\thicksim&(1\pm 0.02997)\\
\end{pmatrix}
\nn
M^{d}=&\chi_{d}\begin{pmatrix}
(0.02102\pm 0.00424)&(0.006014\pm 0.001289)&(0.03921\pm 0.002804)\\
\thicksim&\mathbf{0}&\mathbf{0}\\
\thicksim&\thicksim&(0.9987\pm 0.04818)\\
\end{pmatrix}
\end{align}
\item\begin{align}\label{4zeroB-3}
M^{u}=&\chi_{u}\begin{pmatrix}
(-0.00001043\pm 0.000005618)&(0.002818\pm 0.0004187) &\mathbf{0} \\
\thicksim&(0.9961\pm 0.02997)&(0.05906\pm 0.002507) \\
\thicksim&\thicksim&\mathbf{0}\\
\end{pmatrix}
\nn
M^{d}=&\chi_{d}\begin{pmatrix}
\mathbf{0}&\mathbf{0}&(0.006058\pm 0.001282)e^{i(-180\pm 0.04312)^\circ}\\
\thicksim&(0.9952\pm 0.04808)&(0.06817\pm 0.009958)e^{i(36.54\pm 4.126)^\circ}\\
\thicksim&\thicksim&(-0.01568\pm 0.004455)\\
\end{pmatrix}
\end{align}
\item\begin{align}
M^{u}=&\chi_{u}\begin{pmatrix}
(0.006004\pm 0.01056)&(0.09705\pm 0.05364) &\mathbf{0} \\
\thicksim&(0.9905\pm 0.03207)&(0.002462\pm 0.002079) \\
\thicksim&\thicksim&\mathbf{0}\\
\end{pmatrix}
\nn
M^{d}=&\chi_{d}\begin{pmatrix}
(0.033\pm 0.0138)&(0.0955\pm 0.0801)e^{i(23.8\pm 17.2)^\circ}&(0.00552\pm 0.00151)e^{i(84.1\pm 42.5)^\circ}\\
\thicksim&(0.991\pm 0.0503)&\mathbf{0}\\
\thicksim&\thicksim&\mathbf{0}\\
\end{pmatrix}
\end{align}
\item\begin{align}
M^{u}=&\chi_{u}\begin{pmatrix}
(0.3918\pm 0.02016)&(0.0002621\pm 0.00005475) &(-0.486\pm 0.01299) \\
\thicksim&\mathbf{0}&\mathbf{0} \\
\thicksim&\thicksim&(0.6115\pm 0.02454)\\
\end{pmatrix}
\nn
M^{d}=&\chi_{d}\begin{pmatrix}
(0.3984\pm 0.02144)&\mathbf{0}&(-0.5045\pm 0.02052)e^{i(4.152\pm 0.3504)^\circ}\\
\thicksim&\mathbf{0}&(0.007499\pm 0.000643)e^{i(-95.89\pm 21.92)^\circ}\\
\thicksim&\thicksim&(0.588\pm 0.03201)\\
\end{pmatrix}
\end{align}
\item\begin{align}
M^{u}=&\chi_{u}\begin{pmatrix}
(0.01519\pm 0.009309)&(0.1355\pm 0.03331) &(0.000218\pm 0.00002899) \\
\thicksim&(0.9814\pm 0.03124)&\mathbf{0} \\
\thicksim&\thicksim&\mathbf{0}\\
\end{pmatrix}
\nn
M^{d}=&\chi_{d}\begin{pmatrix}
\mathbf{0}&(0.1574\pm 0.01544)e^{i(-14\pm 5.799)^\circ}&(0.006954\pm 0.001226)e^{i(48.96\pm 18.69)^\circ}\\
\thicksim&(0.9714\pm 0.04805)&\mathbf{0}\\
\thicksim&\thicksim&(-0.0005743\pm 0.0006804)\\
\end{pmatrix}
\end{align}
\item\begin{align}
M^{u}=&\chi_{u}\begin{pmatrix}
(0.9797\pm 0.03049)&(0.1421\pm 0.02397) &(-0.001423\pm 0.0003427) \\
\thicksim&(0.01705\pm 0.007037)&\mathbf{0} \\
\thicksim&\thicksim&\mathbf{0}\\
\end{pmatrix}
\nn
M^{d}=&\chi_{d}\begin{pmatrix}
(0.978\pm 0.0483)&(-0.149\pm 0.0177)e^{i(-164\pm 2.03)^\circ}&\mathbf{0}\\
\thicksim&\mathbf{0}&(0.00669\pm 0.00149)e^{i(18.6\pm 14.62)^\circ}\\
\thicksim&\thicksim&(-0.000666\pm 0.00072)\\
\end{pmatrix}
\end{align}
\end{enumerate}

\bibliographystyle{apsrev}
\bibliography{/home/zhou/reflist/reflist,/home/zhou/reflist/misc}
\end{document}